\documentstyle[12pt]{article}
\title{Probing CP violation with time integrated decay rates into
non-CP eigenstates}
\author{Jo\~ao P.\ Silva\\
\\
\small Centro de F\'{\i}sica Nuclear da Univ. de Lisboa\\
\small Av.\ Prof.\ Gama Pinto, 2,
\small 1699 Lisboa Codex, Portugal\\
\small and \\
\small Centro de F\'{\i}sica, Instituto Superior de Engenharia de Lisboa}
\begin{document}
\maketitle
\begin{abstract}
Many of the experiments proposed to look for interference CP violation
in neutral $B$ mesons concentrate on tagged decay into CP eigenstates.
Aleksan, Dunietz, Kayser and Le Diberder have shown that one can also
look for interference CP violation using tagged decays into non-CP eigenstates.
In all these methods, one must trace the time dependence of the decays.
In this article we discuss a new method to search for interference
CP violation by using only time integrated rates into non-CP eigenstates.
The method hinges on the comparison between the decays of
$\Upsilon(4S)$ into  $f f$, $\bar{f} \bar{f}$, and $f \bar{f}$,
and also uses the rates for $l^+ f$ and $l^- f$.
This method does not depend on how the $\Upsilon(4S)$ is produced;
provided enough statistics,
one can use both symmetric and asymmetric colliders.
\end{abstract}

% \newpage

\section{Introduction}

The standard model explanation of CP violation through a complex phase
in the Cabibbo-Kobayashi-Maskawa matrix \cite{CKM} will be subject to
close scrutiny with the advent of the upcoming $B$-factories.
Since CP violation is an effect of order $10^{-3}$ with respect to
the CP conserving portion of the weak interactions,
it is naturally hard to probe.
In $B$ decays, small CP violating effects are
expected for decay channels with large branching ratios,
and large asymmetries can be found,
but in channels with small branching ratios.
Therefore, one must test CP violation in as many
decay modes as possible.
If one starts with (flavour-) tagged initial states,
a very clean measurement of a CKM CP-violating phase is obtained
by looking for the decay into a CP eigenstate,
whenever this decay is dominated by a single weak phase \cite{Sanda}.
The gold-platted signal is provided by the $B \rightarrow J/\Psi K_S$
rate asymmetry,
enabling a clean measurement of the CKM angle $\beta$.
But, most other CP-eigenstates are considerably more difficult to detect.
Therefore, it is very important to study also non-CP eigenstate 
decay modes.
This has been pointed out by
Aleksan, Dunietz, Kayser and Le Diberder who have shown that CP violation
can also be probed by tracing the time dependence of tagged decays
into non-CP eigenstates \cite{Ale91a}.
Here, we wish to show that one can also extract CKM phases
by relating several $\Upsilon(4S)$ decays into {\em two} non-CP eigenstate
final states, and, moreover,
that this can be achieved through the use of time integrated rates alone.

The Babar and Belle detectors will operate at asymmetric $e^+ e^-$
colliders producing $\Upsilon(4S)$  ($J^{PC}=1^{--}$),
which decays around $50\%$ of the times into a correlated
$B^0 - \overline{B^0}$ pair,
\begin{equation}
| \Phi^- \rangle = \frac{1}{\sqrt{2}}
\left[
| B^0 (\vec k) \rangle \otimes | \overline{B^0} (- \vec k)\rangle
- | \overline{B^0} (\vec k)\rangle \otimes |B^0 (- \vec k) \rangle
\right].
\end{equation}
When the initial $\overline{B^0}$ meson evolves in time,
it oscillates back and forth into $B^0$,
and vice-versa.
But,
the antisymmetric nature of the wavefunction produces a well
known correlation effect between the two mesons.
Indeed, since two identical mesons cannot be in a totally
antisymmetric state, tagging the flavour of one meson through its
semileptonic decay at time $t_2$ ensures that the other meson must
have the opposite flavour at that time $t_2$.
Thenceforth, the surviving meson will evolve as a tagged meson with
time variable $t_- = t_1 - t_2$.
In the standard notation, we have
\begin{eqnarray}
\label{fantasticomike}
\left| \langle f; l^-; t_- | T | \Phi^- \rangle \right|^2
& = & \frac{\left| A_l \right|^2}{4 \Gamma}
\Gamma [ B^0 (t_-) \rightarrow f ],
\nonumber\\
\left| \langle f; l^+; t_- | T | \Phi^- \rangle \right|^2
& = & \frac{\left| A_l \right|^2}{4 \Gamma}
\Gamma [ \overline{B^0} (t_-) \rightarrow f ].
\end{eqnarray}
The substitution of
$|\langle X^- l^+ \nu |T| B^0 \rangle|$
and
$|\langle X^+ l^- \bar \nu |T| \overline{B^0} \rangle|$
by a common value $|A_l|$,
follows from the lack of direct CP violation in semileptonic decays.
In writing the equalities in eqs.~(\ref{fantasticomike})
we have assumed that $t_- >0$.
The expressions for  $t_- < 0$ are found by replacing $t$ by $|t|$ in the
exponentials of the tagged decay rates,
\begin{eqnarray}
\label{master:t}
\frac{\Gamma [B^0 (t) \rightarrow f]}{|A_f|^2}
& = &
\frac{e^{- \Gamma t}}{2}
\left\{
\left( 1 + |\lambda_f|^2 \right)
+
\left( 1 - |\lambda_f|^2 \right) \cos{\Delta m t}
+
2 \mbox{Im} \lambda_f \sin{\Delta m t}
\right\},
\nonumber\\*[2mm]
\frac{\Gamma [\overline{B^0} (t) \rightarrow f]}{|A_f|^2}
& = &
\frac{e^{- \Gamma t}}{2}
\left\{
\left( 1 + |\lambda_f|^2 \right)
-
\left( 1 - |\lambda_f|^2 \right) \cos{\Delta m t}
-
2 \mbox{Im} \lambda_f \sin{\Delta m t}
\right\}.
\end{eqnarray}
We define $A_f = \langle f|T|B^0 \rangle$,
$ \bar{A}_f = \langle f|T|\overline{B^0} \rangle$,
and
$\lambda_f = q/p\, \bar{A}_f/A_f$.
The parameters $q$ and $p$ describe the transformation from
the $B^0, \overline{B^0}$ basis into the basis formed by the eigenstates
of the hamiltonian.

We will assume throughout that there is no CP violation in the mixing,
$|q/p|=1$ (or $q/p= \exp{2 i \phi_m}$),
and that the difference in the lifetimes is negligible,
$\Delta \Gamma = 0$.
Moreover,
we will simplify the discussion by concentrating on decays dominated by a
single weak phase.
In this case, we may write $A_f = A \exp{i(\phi_a + \delta_a)}$,
where $A$ is a magnitude, $\phi_a$ is a weak phase and
$\delta_a$ a strong phase. 
As a consequence,
for the CP-conjugate final state $\bar{f}$ one
gets\footnote{We will discard
all phases brought about by CP transformations.}
$\bar{A}_{\bar f} = A \exp{i(- \phi_a + \delta_a)}$,
$|A_f|=|\bar{A}_{\bar f}|$,
and there is no direct CP violation.

Most authors refer precisely to the experiments described by
eq.~(\ref{fantasticomike}). One tags one meson through its semileptonic
decay and looks for the decay of the other meson into some final state $f$.
When $f$ is a CP eigenstate, and under the assumptions just described,
$|\lambda_f|=1$ and
$\mbox{Im} \lambda_f = \pm \sin 2 (\phi_m - \phi_a)$.
The standard CP-violating asymmetry becomes \cite{Sanda}
\begin{eqnarray}
\label{A(t-)}
A(t_-) &=&
\frac{\left| \langle f_{cp}; l^-; t_- | T | \Phi^- \rangle \right|^2
-\left| \langle f_{cp}; l^+; t_- | T | \Phi^- \rangle \right|^2
}{\left| \langle f_{cp}; l^-; t_- | T | \Phi^- \rangle \right|^2
+\left| \langle f_{cp}; l^+; t_- | T | \Phi^- \rangle \right|^2
}
\nonumber\\*[2mm]
&=&
\mbox{Im} \lambda_f\ \sin{\Delta m t_-},
\end{eqnarray}
thus measuring a CKM phase $\phi_m - \phi_a$ directly.
When $f$ is not a CP eigenstate,
one needs two sets of parameters to describe the amplitudes,
\begin{eqnarray}
\label{asamplitudes}
A_f = A e^{i \phi_a} e^{i \delta_a},
& \hspace{5mm} & 
\bar{A}_{\bar f} = A e^{-i \phi_a} e^{i \delta_a},
\nonumber\\*[1mm]
\bar{A}_f = B e^{i \phi_b} e^{i \delta_b},
& \hspace{5mm} &
A_{\bar f} = B e^{-i \phi_b} e^{i \delta_b}.
\end{eqnarray}
Therefore
\begin{equation}
\label{oslambdas}
\lambda_f = \frac{B}{A} e^{2 i \phi} e^{i \Delta},
\hspace{5mm}
\lambda_{\bar f} = \frac{A}{B} e^{2 i \phi} e^{- i \Delta},
\end{equation}
where $2 \phi \equiv 2 \phi_m + \phi_b - \phi_a$
and $\Delta \equiv \delta_b - \delta_a$.
In reference \cite{Ale91a},
it is shown that by comparing the four {\em tagged, time-dependent}
decay rates of $\Upsilon(4S)$ into
$f l^-$, $f l^+$, $\bar{f} l^-$ and $\bar{f} l^+$,
one can recover $B/A$, $\sin(2 \phi + \Delta)$,
$\sin(2 \phi - \Delta)$, and the normalization factor
$A^2 + B^2$. This yields the weak phase $\phi$, up to discrete ambiguities.

\section{Time integrated rates}

It is very important to note that the interference CP violation terms,
which are those proportional to $\mbox{Im} \lambda_f$,
appear in the {\em tagged} decay rates, 
multiplied by the function $e^{- \Gamma |t_-|} \sin{\Delta m t_-}$.
This function is odd under $t_1 \leftrightarrow t_2$ and, hence,
it vanishes when we compute the time integrated rates,
from $t_-= - \infty$ to $t_-= + \infty$.
This feature is common to all methods based on the tagged decays of
the $\Upsilon(4S)$:
one must follow the time dependence (or, at least, the time ordering)
in order to measure interference CP violation.

Of course,
this does not occur for the time integrated rate of a single meson,
since, then, the individual decay times,
$t_1$ and $t_2$, run from $0$ to $+ \infty$.
In fact,
\begin{eqnarray}
\label{master}
\Gamma_f \equiv 
\Gamma [B^0 \rightarrow f]
& = &
\frac{|A_f|^2}{2 \Gamma} 
\left\{
\left( 1 + |\lambda_f|^2 \right)
+
\frac{
\left( 1 - |\lambda_f|^2 \right)
+
2 x \mbox{Im} \lambda_f}
{1 + x^2}
\right\},
\nonumber\\*[2mm]
\bar{\Gamma}_f \equiv
\Gamma [\overline{B^0} \rightarrow f]
& = &
\frac{|A_f|^2}{2 \Gamma}
\left\{
\left( 1 + |\lambda_f|^2 \right)
- \frac{
\left( 1 - |\lambda_f|^2 \right) 
+
2 x \mbox{Im} \lambda_f}{1+x^2}
\right\},
\end{eqnarray}
where $x= \Delta m /\Gamma \sim 0.73$ and we recall that
$|q/p|=1$ is assumed throughout.

The time integrated rate of the initial two meson state into a final state
$f$ and a final state $g$ is proportional to
\begin{equation}
\label{integrated}
R_{fg} \equiv
\int_0^{+ \infty} dt_1\, 
\int_0^{+ \infty} dt_2\,
\left| \langle f,t_1;g,t_2 | T | \Phi^- \rangle \right|^2
=
\frac{1}{2}
\left[
\Gamma_f \bar{\Gamma}_g + \Gamma_g \bar{\Gamma}_f
- 2 \mbox{Re} \left( I_f I_g^\ast \right)
\right].
\end{equation}
The parameters $I_f$ describe the interference term and have been derived
in the appendix. We find
\begin{equation}
\label{Ifunctions}
I_f =
\frac{|A_f|^2}{2 \Gamma}
\left\{
2 \mbox{Re} \lambda_f
- i
\frac{2 \mbox{Im} \lambda_f - 
\left( 1 - |\lambda_f|^2 \right) x}{1 + x^2}
\right\}.
\end{equation}

\section{Decays into non-CP eigenstates}

Let us now compare the three time integrated decay rates into
$f f$, $f \bar{f}$ and $\bar{f} \bar{f}$.
Using eqs.~(\ref{master}) through (\ref{Ifunctions}),
we find
\begin{eqnarray}
\label{Rff}
R_{ff} & = &
N\, \left| \frac{A_f}{A_{\bar f}} \right|^2\,
\left| \lambda_f^2 - 1 \right|^2,
\\*[2mm]
\label{Rfbarfbar}
R_{\bar{f} \bar{f}} & = &
N\, \left| \frac{A_{\bar f}}{A_f} \right|^2\,
\left| \lambda_{\bar f}^2 - 1 \right|^2,
\\*[2mm]
\label{Rffbar}
R_{ f \bar{f}} \equiv R_{\bar{f} f}  & = &
N\, \left[
\left| \lambda_f \lambda_{\bar f} - 1 \right|^2
+ \frac{x^2+2}{x^2} \left| \lambda_f - \lambda_{\bar f} \right|^2
\right],
\end{eqnarray}
where
\begin{equation}
N = \frac{|A_f|^2 |A_{\bar f}|^2}{4 \Gamma^2}
\frac{x^2}{1+x^2}
\end{equation}
is a common normalization factor.

When the decays are dominated by a single weak phase,
eqs.~(\ref{asamplitudes}) and (\ref{oslambdas}) hold.
Substituting in eqs.~(\ref{Rff}) -- (\ref{Rffbar}), one gets,
\begin{eqnarray}
\label{hello}
R_{ff} & \propto &
\frac{A^2}{B^2} + \frac{B^2}{A^2}
- 2 \cos{\left( 4 \phi + 2 \Delta \right)}
\nonumber\\*[2mm]
R_{\bar{f} \bar{f}} & \propto &
\frac{A^2}{B^2} + \frac{B^2}{A^2}
- 2 \cos{\left( 4 \phi - 2 \Delta \right)}
\nonumber\\*[2mm]
R_{ f \bar{f}} \equiv R_{\bar{f} f}  & \propto &
4 \sin^2 (2 \phi)
+ \frac{x^2+2}{x^2} 
\left[ 
\frac{A^2}{B^2} + \frac{B^2}{A^2}
- 2 \cos{(2 \Delta)}
\right].
\end{eqnarray}
We have three rates but four quantities to be fitted:
$\phi$, $\Delta$, $A/B$, and the overall normalization $N$.
The normalization can be fitted in a variety of ways.
For example, we can integrate eqs.~(\ref{fantasticomike})
making use of eqs.~(\ref{master}).
We merely have to note that, since the time integration is now from 
$t_- =- \infty$ to $t_- = + \infty$,
the expressions (\ref{master}) get multiplied by a factor of two and the
$\mbox{Im} \lambda_f$ terms disappear,
\begin{eqnarray}
R_{f l^-} \equiv R_{\bar{f} l^+} & = &
\frac{|A_l|^2}{4 \Gamma^2}
\left[
A^2 + B^2 + \frac{A^2 - B^2}{1+x^2}
\right],
\nonumber\\*[2mm]
R_{f l^+} \equiv R_{\bar{f} l^-} & = &
\frac{|A_l|^2}{4 \Gamma^2}
\left[
A^2 + B^2 - \frac{A^2 - B^2}{1+x^2}
\right],
\end{eqnarray}
These rates give us $A$ and $B$.
Therefore, we can extract $\phi$ and $\Delta$ from
eqs.~(\ref{hello}), up to discrete ambiguities,
and we still get a consistency equation for free.
This can be used to remove some of the discrete ambiguities.

Some examples of relevant non-CP eigenstates are $D^{\ast +} D^-$ and
$\rho^+ \pi^-$ \cite{Ale91a}.
The former measures the angle $\beta$;
the latter measures the angle $\alpha = \pi -(\beta + \gamma)$
and was discussed in detail in reference \cite{Ale91a}.
In addition,
the decays into $D^0 K_S$ are very interesting because
they provide a measurement of the unusual combination $(2 \beta + \gamma)$
\cite{Gro91}.
A general problem with all final states common to $B^0_d$ and
$\overline{B^0_d}$,
is that the respective branching ratios are not expected to 
exceed $10^{-3}$.
The exception occurs for $B^0_d \rightarrow D^- \pi^+$,
with a branching of $(3.0 \pm 0.4) \times 10^{-3}$
\cite{PDG}.
However,
we also need $B^0_d \rightarrow D^+ \pi^-$ whose amplitude is further
doubly Cabibbo suppressed.
The situation is much better for the $B^0_s - \overline{B^0_s}$
pairs \cite{Ale91b},
but these will be produced in the near future only in uncorrelated states.

One can also consider decays into two CP eigenstates,
$\Upsilon(4S) \rightarrow f_{cp} f_{cp}$. In this case,
$\lambda_f = \lambda_{\bar f}$
and eqs.~(\ref{Rff}) -- (\ref{Rffbar}) reduce to
\begin{equation}
\label{Rfcpfcp}
R_{f_{cp} f_{cp}} = N\, \left| \lambda_f^2 - 1 \right|^2
= N\,
\left[ 
  \left(|\lambda_f|^2 - 1\right)^2
  + 4 \mbox{Im}^2 \lambda_f
\right].
\end{equation}
When the decay is dominated by a single weak phase this is a clean
observable of $\sin^2(2 \phi)$ \cite{fcpfcp}.
We should add the following observation.
When these tests become available,
one can measure interference CP violation even in the presence
of direct CP violation (and in either symmetric or asymmetric colliders).
In fact,
although the interference CP violation term (proportional to
$\sin{\Delta m t_-}$) cancels when we integrate the asymmetry
(\ref{A(t-)}) over time,
the direct CP violation term (proportional to $\cos{\Delta m t_-}$)
does not.
The time integrated tagged asymmetry yields a measurement
of $|A_f|^2 - |\bar{A}_f|^2$ \cite{Des96}.
Comparing this with (\ref{Rfcpfcp}),
we can determine $\mbox{Im}^2 \lambda_f$,
even when there is also direct CP violation.
Unfortunately,
in this case $\mbox{Im} \lambda_f$ does not provide a clean measurement
of a CKM phase since $\lambda_f$ will involve the two complex
amplitudes contributing to the decay into the CP eigenstate
\cite{Gro93}.

\section{Conclusions}

In this article,
we build on previous work by 
Aleksan, Dunietz, Kayser and Le Diberder,
who have considered tagged, time-dependent decays of
neutral $B$ mesons into non-CP eigenstates.
We show that one can determine CKM CP-violating
phases by using exclusively {\em time integrated} decay rates of
neutral $B$ mesons into non-CP eigenstates,
if the decays are dominated by a single weak phase.
Naturally, such experiments can be performed either at asymmetric or at
symmetric $e^+ e^-$ colliders.
Our result is one further violation of the frequent misconception that
symmetric $b$-factories cannot, even in principle,
be used to measure interference CP violation.
In fact, it has long be known that one can detect $\mbox{Im} \lambda_f$
in the decays into two CP eigenstate final states \cite{fcpfcp}.

The best channel for our analysis is probably $D^{\ast +} D^-$.
The most interesting is $D^0 K_S$,
for it measures the angle combination $(2 \beta + \gamma)$.
However,
the branching ratios are such that one must wait for second generation 
experiments.
This is a standard feature of decays into two states,
both common to $B^0_d$ and $\overline{B^0_d}$.
The situation improves considerably for the $B^0_s$ system.

%%%%%%%%%%%%%%%%%%%%%

\vspace{5mm}

I am indebted to A.\ Barroso and L.\ Lavoura for many discussions,
and to L.\ Wolfenstein for very elucidating remarks on this subject.
I thank them for reading and criticizing the manuscript.
This work was supported by the Portuguese JNICT under contract
CERN/P/FIS/1096/96.

\vspace{5mm}

%%%%%%%%%%%%%%%%%%%%%%
\appendix

\section{Appendix}

In the neutral $B$ system,
the mass eigenstates may be written in terms of the flavour eigenstates as
\begin{eqnarray}
|B_H \rangle &=& p |B^0 \rangle + q |\overline{B^0} \rangle,
\nonumber\\
|B_L \rangle &=& p |B^0 \rangle - q |\overline{B^0} \rangle.
\end{eqnarray}
Therefore, a flavour eigenstate created at time $t=0$ will
evolve in time as
\begin{eqnarray}
|B^0(t) \rangle &=& g_+(t) |B^0 \rangle
+ \frac{q}{p} g_-(t)|\overline{B^0} \rangle,
\nonumber\\
|\overline{B^0}(t) \rangle &=& \frac{p}{q} g_-(t)|B^0 \rangle 
+ g_+(t) |\overline{B^0} \rangle.
\end{eqnarray}
where
\begin{equation}
g_\pm = \frac{1}{2} \left( e^{- i \mu_H t} \pm e^{- i \mu_L t}\right),
\end{equation}
and $\mu_\alpha = m_\alpha - i \Gamma_\alpha / 2$
($\alpha = H, L$) are the eigenvalues of the mixing matrix. 
The amplitude for the neutral meson to decay into a final state $f$
at time $t$, is given by
\begin{eqnarray}
\label{aux1}
\langle f|T|B^0(t) \rangle &=& 
A_f \left[ g_+(t)  + \lambda_f g_-(t) \right],
\nonumber\\
\langle f|T|\overline{B^0}(t) \rangle &=& 
\frac{p}{q} A_f \left[ g_-(t)  + \lambda_f g_+(t) \right],
\end{eqnarray}
where $A_f = \langle f|T|B^0 \rangle$,
$ \bar{A}_f = \langle f|T|\overline{B^0} \rangle$,
and
$\lambda_f = q/p\, \bar{A}_f/A_f$.
Eqs.~(\ref{master:t}) follow trivially from these,
when one uses the definitions
$\Gamma = (\Gamma_H + \Gamma_L)/2$,
$\Delta \Gamma = \Gamma_H - \Gamma_L$,
$m = (m_H + m_L)/2$,
and $\Delta m = m_H - m_L$ ($\Delta m$ is positive by convention).

To find the time integrated rates we need
\begin{eqnarray}
\label{aux2}
G_\pm \equiv \int_0^\infty dt\ |g_\pm(t)|^2 
&=& \frac{1}{2 \Gamma}
\left(
\frac{1}{1-y^2} \pm \frac{1}{1+x^2}
\right),
\nonumber\\
G_{+-} \equiv \int_0^\infty dt\ g_+^\ast(t) g_-(t) 
&=& \frac{1}{2 \Gamma}
\left(
\frac{-y}{1-y^2} + \frac{-ix}{1+x^2}
\right),
\end{eqnarray}
where $y = \Delta \Gamma/(2 \Gamma)$,
and $ x = \Delta m/\Gamma$.
From eqs.~(\ref{aux1}) and (\ref{aux2}) we may derive the time integrated
rates in eqs.~(\ref{master}).

Finally, for eq.~(\ref{integrated}) we need the integrals
\begin{eqnarray}
\label{aux3}
I_\psi &\equiv& \int_0^\infty dt\ \langle \psi|T|B^0 \rangle
{\langle \psi|T|\overline{B^0} \rangle}^\ast
\nonumber\\
&=&
\frac{p^\ast}{q^\ast} |A_\psi|^2
\left\{
\mbox{Re} \lambda_\psi \left(G_+ + G_- \right) +
\left( 1 + |\lambda_\psi|^2 \right) \mbox{Re} G_{+-}
\right.
\nonumber\\
& &
\hspace{15mm}
\left.
- i \mbox{Im} \lambda_\psi \left(G_+ - G_- \right) 
- i\left( 1 - |\lambda_\psi|^2 \right) \mbox{Im} G_{+-}
\right\}
\nonumber\\*[2mm]
& = &
\frac{p^\ast}{q^\ast} \frac{|A_\psi|^2}{2 \Gamma}
\left\{
\frac{2 \mbox{Re} \lambda_\psi - \left( 1 + |\lambda_\psi|^2 \right)\, y}{
1-y^2}
%\right.
%\nonumber\\
%& &
%\hspace{5mm}
%\left.
- i \frac{2 \mbox{Im} \lambda_\psi - \left( 1 - |\lambda_\psi|^2 \right)\, x}{
1+x^2}
\right\}.
\end{eqnarray}
Eq.~(\ref{Ifunctions}) follows from this when we take $y=0$,
if we note that, under the assumption that $|q/p|=1$,
the $p/q$ prefactor in eqs.(\ref{aux3}) cancels in the $I_f I_g^\ast$
product relevant for eq.~(\ref{integrated}).

% \newpage

%
 

\begin{thebibliography}{99}
%
\bibitem{CKM}
N.\ Cabibbo,
Phys.\ Rev.\ Lett.\ {\bf 10}, 531 (1963);
M.\ Kobayashi and T.\ Maskawa,
Prog.\ Theor.\ Phys.\ {\bf 49}, 652 (1973).
%
\bibitem{Sanda}
A.\ B.\ Carter and A.\ I.\ Sanda,
Phys.\ Rev.\ Lett.\ {\bf 45}, 952 (1980);
Phys.\ Rev.\ D {\bf 23}, 1567 (1981);
I.\ I.\ Bigi and A.\ I.\ Sanda,
Nucl.\ Phys.\  {\bf B193}, 85 (1981);
{\bf B281}, 41 (1987).
%
\bibitem{Ale91a}
R.\ Aleksan, I.\ Dunietz, B.\ Kayser, and F.\ Le Diberder,
Nucl.\ Phys.\  {\bf B361}, 141 (1991).
%
\bibitem{Gro91}
M.\ Gronau and D.\ London,
Phys.\ Lett.\ B {\bf 253}, 483 (1991).
%
\bibitem{PDG}
Particle Data Group,
L.\ Montanet {\it et al.},
Phys.\ Rev.\ D {\bf 50}, 1173 (1994).
%
\bibitem{Ale91b}
See, for example,
R.\ Aleksan, I.\ Dunietz, and B.\ Kayser,
Z.\ Phys.\ C {\bf 54}, 653 (1992).
%
\bibitem{fcpfcp}
L.\ Wolfenstein,
Nucl.\ Phys.\  {\bf B246}, 45 (1984);
M.\ B.\ Gavela {\it et. al.},
Phys.\ Lett.\ B {\bf 162}, 197 (1985);
I.\ I.\ Bigi and A.\ I.\ Sanda,
(1987), in reference \cite{Sanda}.
%
\bibitem{Des96}
N.\ G.\ Deshpande and X.~-G.\ He,
Phys.\ Rev.\ Lett.\ {\bf 76}, 360 (1996).
%
\bibitem{Gro93}
M.\ Gronau,
Phys.\ Lett.\ B {\bf 300}, 163 (1993).
%
\end{thebibliography}
\end{document}